# Accuracy, Repeatability, and Reproducibility of Firearm Comparisons

# Part 1: Accuracy

L. Scott Chumbley, Max D. Morris, Stanley J. Bajic, and Daniel Zamzow

Ames Laboratory-USDOE

Erich Smith, Keith Monson, Gene Peters*

Federal Bureau of Investigation

**Abstract:**  Researchers at the Ames Laboratory-USDOE and the Federal Bureau of Investigation (FBI) conducted a study to assess the performance of forensic examiners in firearm investigations.  The study involved three different types of firearms and 173 volunteers who compared both bullets and cartridge cases.  The total number of comparisons reported is 20,130, allocated to assess accuracy (8,640), repeatability (5,700), and reproducibility (5,790) of the evaluations made by participating examiners. The overall false positive error rate was estimated as 0.656% and 0.933% for bullets and cartridge cases, respectively, while the rate of false negatives was estimated as 2.87% and 1.87% for bullets and cartridge cases, respectively. Because chi-square tests of independence strongly suggest that error probabilities are not the same for each examiner, these are maximum likelihood estimates based on the beta-binomial probability model and do not depend on an assumption of equal examiner-specific error rates. Corresponding 95% confidence intervals are (0.305%,1.42%) and (0.548%,1.57%) for false positives for bullets and cartridge cases, respectively, and (1.89%,4.26%) and (1.16%,2.99%) for false negatives for bullets and cartridge cases, respectively. These results are based on data representing all controlled conditions considered, including different firearm manufacturers, sequence of manufacture, and firing separation between unknown and known comparison specimens. The results are consistent with those of prior studies, despite its more robust design and challenging specimens.

# Introduction

Forensic firearms examination, like other pattern evidence analysis disciplines (e.g., latent fingerprints, LP), relies on expertise, training, and judgment to make comparisons between questioned, evidentiary specimens and known exemplars for source attribution decisions. The Federal Bureau of Investigation (FBI) Laboratory initiated research to strengthen the admissibility of pattern evidence examination decisions in 2006, starting initially with fingerprint comparisons, publishing the first results in 2011 [1]. This paper is the first in a series that will report results obtained concerning the accuracy, repeatability, reproducibility, and other factors influencing firearm examinations.

**Previous Studies**

Previous studies by forensic firearm examiners and independent researchers [2-7] have also examined the accuracy of firearm examiner decisions. The President's Council of Advisors on Science and Technology (PCAST) published a 2016 report titled "Forensic Science in Criminal Courts: Ensuring Scientific Validity of Feature-Comparison Methods." The PCAST report [8] reviewed the scientific validity of a number of feature-comparison analysis methods, including latent fingerprints (LP) and firearms. Concerning fingerprints, PCAST concluded that the design of and results from an LP study [1] were instrumental in establishing the validity of LP comparisons. PCAST also reviewed a Department of Energy (DoE) report in the public domain by Baldwin *et al*. [7] in the area of firearm examinations. The experimental methodology used in both these studies, particularly the open set design were described by PCAST to be of high quality. The praise given these studies by PCAST affirmed the experimental design approach used in this study [9].

Since publication of the PCAST report a number of firearms examiners and independent researchers have conducted additional investigations dealing with various aspects of comparative examinations. These include efforts to produce either automated or computer-based objective determinations [11-17]; statistical evaluation methods in the identification of toolmarks [18,19]; and estimation of examiner error rates [20-24]. Additionally, several review compilations contain general discussions as applied to firearms and toolmark examinations [25,26].

**Study Design**

The study discussed in this paper was designed to provide statistically valid information concerning examiner accuracy, repeatability, and reproducibility [9]. It was designed as a true double-blind "black box" investigation, with contact between the participating examiner subjects and the experimental team restricted at all times to both preserve anonymity of the participants and prevent any interactions between participants and investigators that might result in bias. Duties related to communication with the participants and generation and scoring of the specimens provided for examination were strictly compartmentalized. No specimen-specific information was shared between the communication group and the experimental / analysis group. All results were kept anonymous and the study was subject to review and approval by the Institutional Review Boards of Ames Laboratory's contracting agency, Iowa State University, and the FBI.

The basic task of the study was the comparison of unknown cartridge cases and bullet specimens by examiner subjects who volunteered to participate. Both bullets and cartridge cases were obtained using three different brands of firearms and a single standard type of ammunition. The firearms and ammunition selected for this study were selected for their propensity to produce challenging and ambiguous test specimens, creating difficult comparisons for examiners [9]. Thus, the study was designed to be a rigorous trial of examiner ability; as a result, error rates derived from this study may provide an upper-bound to operational casework, as evidentiary specimens may be generally assumed to be less challenging than those used in this study. Specimens were provided to the volunteers through a series of mailings and participants were specifically asked not to use their laboratory or agency peer-review process and not to discuss their conclusions with others. Analogous to the previous LP [1] and Baldwin [7] studies, in terms of experimental design and methodology, this study was broader than Baldwin in that it involved both cartridge cases and bullets and took into account additional parameters that might affect examiner accuracy such as challenging comparisons, manufacturing conditions, presence of subclass characteristics, and firing order for specimen comparisons.

Broad calls for volunteers were made through the Association of Firearm and Toolmark Examiners (AFTE) website, by announcements by FBI personnel at national forensic meetings, through e-mail lists maintained by AFTE, and through national / international listservs. Due to difficulties with mailing bullets / cartridge cases overseas, a decision was made to accept only examiners within the United States. Examiners associated with the FBI were also excluded to eliminate possible conflicts of interest. Some initial volunteers discontinued their participation in the project without reviewing any specimen packets once it was realized that their daily workload was not compatible with the amount of effort required to complete the study. A total of 173 examiners returned evaluations and were active participants in the study.

This paper reports study results related to examiner accuracy (error rates). An accompanying paper (Part 2) covers the repeatability and reproducibility of examination results. Future papers in this series will elaborate on the experimental design and will examine effects related to firearms, tool wear, and human factors.

## **Experimental Procedure**

Details of experimental design, pilot testing, specimen generation, rationale for choosing particular firearms to produce bullets or cartridge cases, and ammunition choice are provided in [Design paper -9]. The ammunition used in this study was Wolf Polyformance 9mm Luger (9x19mm). The cartridges were steel case 115 grain full metal jacket (FMJ) rounds. Cartridge cases were collected from 11 Jimenez and 27 Beretta semiautomatic handguns. Bullets were collected from 11 Ruger and the same 27 Beretta semiautomatic handguns. The number of specimens collected was 700 specimens per Beretta firearm and 850 specimens per non-Beretta firearm, for both cartridge cases and bullets, from 28,250 test fires. The ammunition was fired and collected by FBI personnel at the FBI Laboratory in Quantico, VA and transferred to Ames Laboratory for distribution to examiners. The majority of the handguns used (43 of the 49) employed newly-, consecutively-manufactured barrels and slides. Each firearm was test-fired to "break it in" before specimen collection began. The break-in firings were necessary to stabilize internal wear within the firearms and achieve consistent and reproducible toolmarks. All firearms were cleaned after each 250 rounds fired during the collection process.

Bullets fired from barrels produced using the same machining tool can result in the introduction of common sub-class characteristics that can complicate comparison [27]. A broaching process produced the firearm barrels used in this study and a typical machining tool may last for the manufacture of a hundred or more barrels. Therefore, barrels machined at different intervals over the lifetime of the broaching tool were collected, so data related to this effect could be obtained; this information will be discussed in a future paper.

Two-dimensional bar codes were used to track all bullets and cartridge cases, as shown in Figure 1. As the study planned to assess repeatability as well as accuracy, these codes made it difficult for participants to identify previously viewed specimens.  Specimens (both cartridge cases and bullets) were labeled as Ks to designate known specimens and labels that had no extra letter were used for questioned (Q) specimens.  This was done so that if an examiner mixed up the known and questioned specimens in a comparison set during analysis, they could still be distinguished. Bullets were epoxied on plastic mounts to facilitate handling and provide a place for affixing labels (Fig. 1b).  Cartridge case labels were placed on areas where minimal marks were present. After all the specimens were labeled, they were inventoried by barcode labels, which were linked to additional information associated with the specimen: serial number of the firearm, test firing-order range, specimen type, and whether the specimen was a questioned or known specimen.  This information was used to track specimens, verify the "ground truth" for all test packets assembled, and for those repackaged for subsequent re-examination in the assessment of examiner reliability and reproducibility.

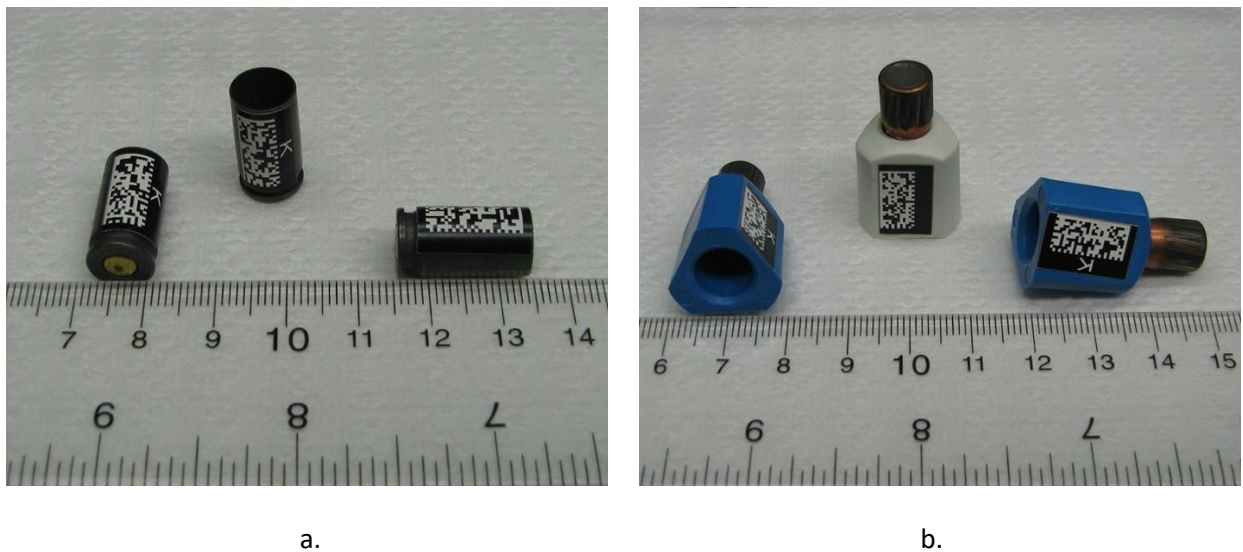

a.          b.

Figure 1: Specimens for examination. a) Cartridge cases showing 2D bar code. b) Bullets showing plastic mounts and 2D bar codes.

Each test packet mailed to examiners consisted of 30 comparison specimen sets; 15 cartridge case sets and 15 bullet sets. Each comparison set consisted of a single questioned specimen to be compared to two known specimens, the latter fired from the same firearm. The cartridge case comparisons were 5 sets of Jimenez and 10 sets of Beretta specimens and the bullet comparisons were 5 sets of Ruger and 10 sets of Beretta specimens.

Test packets and comparison sets were assembled using the following parameters:

1) Only cartridge cases and bullets fired from the same make and model firearm were compared.

2) Each set represented an independent comparison, unrelated to any other set in the test packet.

3) An open-set design was used, i.e., there was not a match for every questioned specimen.

4) The overall proportion of known (true) matches in the test packets was approximately 33%, but varied between bullets and cartridge cases within a test packet and across all test packets.

5) The ratio of non-Beretta to Beretta specimens (for cartridge cases and bullets) in a test packet was 1:2.

Individual cartridge cases and bullets were also divided into three groups, designating the firing order of the collected specimens as early (E), middle (M), or late (L). The firing order was not disclosed to participating subjects, but tracked to evaluate the effect of firearm wear on examiners' analysis results. Results related to firearm manufacture and wear will be discussed in a future paper.

The test packets were distributed through a series of mailings, with examiners being asked to complete as many mailings as possible (up to a total of six) as their time and workload allowed. Each test packet assembled for mailing consisted of the 15 bullet and 15 cartridge case comparison sets to be analyzed, an instruction sheet, and answer sheets [9]. A 3-page survey form was included in the first mailing to obtain demographic and laboratory procedure information, and summarized survey information will be presented in [37]. Each test packet was assigned a primary numeric code to track the comparison sets. The specimen set pairings within a given packet were maintained throughout the study.

Items for each mailing were placed into a Tyvek envelope, sealed, and labeled with a secondary identifying code linked to the unique primary identifying group number for a given test packet. The secondary code was a 3-digit alphanumeric identifier – the first two letters designated the participant while the third letter or number indicated the mailing. This secondary code is what enabled packets to be shared between the experimental group and the communication group; the primary identifiers were not made known to the communication group. A return Tyvek envelope, also labeled with the secondary-identifier code, was included in the assembled packet and used by examiners to return the specimens and their analysis results sheets. Figure 2 shows the contents of a typical first mailing.

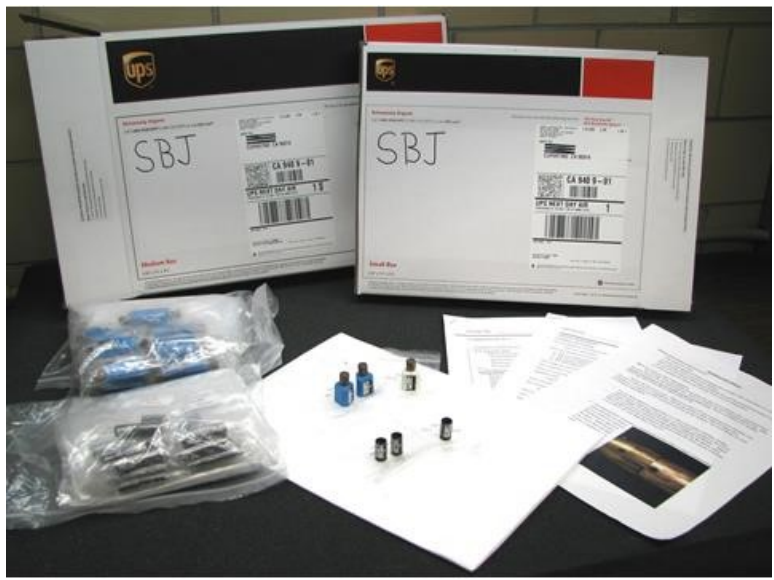

Figure 2: Specimen packet components, including the bullet and cartridge case test sets, forms, and shipping and return boxes.

The sealed Tyvek envelope was transferred to the communication group, who placed the envelope in a shipping box and mailed it to the participating examiner, whose identity was known only to the communications group. Returned packets were logged and inspected upon arrival by the communication group for any examiner-specific identifying information, prior to transferring the sealed Tyvek bag containing the analysis results to the experimental/analysis group for scoring, database entry, and verification of the results.

If a decision error was noted, the comparison set was barcode-read and the information compared to the known "ground truth" to verify the error. Analyzed test packets needed for the assessment of repeatability and reproducibility were repackaged before redistribution.  Each specimen in each comparison set was visually examined and gently cleaned of any debris or marks and assigned new randomized set and group numbers.  The repackaged test packet was barcode-read once again, to verify the "ground-truth" of the reassembled packet, prior to use in succeeding mailings.

The 173 participating firearm examiners provided analysis results for a total of 668 test packets, comprising cartridge cases and bullets.  The total number of specimen-set comparisons reported is 20,130.  Slightly more cartridge case (10,110) than bullet (10,020) comparison are reported because some examiners returned partially-completed test packets that had results for all the cartridge case sets but not all the bullet sets.

Examiners were asked to render a decision for each individual comparison set analyzed as either *Identification*, *Elimination*, *Inconclusive,* or *Unsuitable* using the AFTE 5-point Range of Conclusions (Figure 3) [28].

> 1. Identification
>    Agreement of a combination of individual characteristics and all discernible class characteristics where the extent of agreement exceeds that which can occur in the comparison of toolmarks made by different tools and is consistent with the agreement demonstrated by toolmarks known to have been produced by the same tool.
> 2. Inconclusive
>    a. Some agreement of individual characteristics and all discernible class characteristics, but insufficient for an identification.
>    b. Agreement of all discernible class characteristics without agreement or disagreement of individual characteristics due to an absence, insufficiency, or lack of reproducibility.
>    c. Agreement of all discernible class characteristics and disagreement of individual characteristics, but insufficient for an elimination.
> 3. Elimination
>    Significant disagreement of discernible class characteristics and/or individual characteristics.
> 4. Unsuitable for examination.

Figure 3.  AFTE Range of Conclusions [28]

The numbers of bullet and cartridge case comparison sets analyzed in each of the three rounds of the study are shown in Table 1. Results from the second and third rounds were used for analysis of examiner repeatability and reproducibility, and will be discussed in a future paper [29].  The analyses reported here are based on only the first of the three subsets of the data, defined as follows:

1) **Accuracy** (the focus of this paper) is defined as the ability of an examiner to correctly identify a known match or eliminate a known nonmatch. The data used for this analysis includes <u>only</u> the 4,320 individual comparison sets that were analyzed by the 173 examiners in the first round of the study.

2) **Repeatability** is defined as the ability of an examiner, when confronted with the exact same comparison once again, to make the same decision as when first examined. The data used for this analysis includes <u>only</u> the 2,835 individual bullet and 2,865 individual cartridge case comparison sets that were analyzed twice by the same examiner in the first and second rounds of the study. The repeatability portion of the study involved 105 examiners.

3) **Reproducibility** is defined as the ability of a second examiner to evaluate a packet previously viewed by a different examiner and reach the same conclusion. The data used for this analysis includes <u>only</u> the 2,865 individual bullet and 2,925 individual cartridge case comparison sets that were analyzed twice, by different examiners in the first and third rounds of the study. The reproducibility portion of the study involved 191 of193 pairs of examiners (with some individual examiners participating in more than one reproducibility pair).

Table 1: Numbers of bullet and cartridge case sets analyzed in each of the three rounds of the study

| | Comparison Sets Analyzed by Round | | | |
|---|---|---|---|---|
| | Round 1 | Round 2 | Round 3 | Total |
| **Bullet** | 4,320 | 2,835 | 2,865 | 10,020 |
| **Cartridge Case** | 4,320 | 2,865 | 2,925 | 10,110 |

## Experimental Results: Accuracy

A summary of the Round 1 evaluations for each of the 4,320 bullet and cartridge case comparisons used to determine accuracy, by actual (true) status of each set, is given in Table 2.

Table 2: First-round bullet and cartridge case summary counts

| | ID | Inconclusive-A | Inconclusive-B | Inconclusive-C | Elimination | Other |
|---|---|---|---|---|---|---|
| **Bullet Evaluations by Set Type** | | | | | | |
| **Matching** | 1,076 | 127 | 125 | 36 | **41** | 24 |
| **Nonmatching** | **20** | 268 | 848 | 745 | 961 | 49 |
| **Cartridge Case Evaluations by Set Type** | | | | | | |
| **Matching** | 1,056 | 177 | 140 | 22 | **25** | 25 |
| **Nonmatching** | **26** | 177 | 637 | 620 | 1,375 | 40 |

The heading "ID" indicates an examiner made an Identification evaluation; Inconclusive-A,B,C evaluations correspond to the inconclusive determinations from the AFTE Range of Conclusions (Figure 3). Counts of hard errors are highlighted in bold. Throughout this paper, a <u>hard error</u> is defined as an instance in which Elimination was declared for a true matching set, or Identification was declared for a true nonmatching set. The final column labeled "Other" in Table 2 includes records for which an evaluation was not recorded, was recorded as Inconclusive without a level designation (A, B, or C), where multiple levels were recorded, or for which the examiner indicated that the material was Unsuitable (defined in Figure 3) for evaluation. Counts recorded in the "other" category are not included in this discussion of accuracy. Summary conclusion percentages are computed by dividing each of the entries in Table 2 by its corresponding row sum, excluding sets classified as "other", and are presented in Table 3. Hence, for example, the proportion of incorrect identifications among nonmatching bullet sets (or false positives, F-Pos) is:

F-Pos = 100% x Identification / (Identification + Inconclusive-A + Inconclusive-B + Inconclusive-C + Elimination)

= 100% x 20 / (20 + 268 + 848 + 745 + 961) = 0.704%

and the proportion of Eliminations among matching bullet sets (or false negatives, F-Neg) is

F-Neg = 100% x Elimination / (Identification + Inconclusive-A + Inconclusive-B + Inconclusive-C + Elimination)

= 100% x 41 / (1076 + 127 + 125 + 36 + 41) = 2.92%

after removal of the comparisons represented in the "other" column of Table 2.

Table 3: First-round summary percentages of bullet and cartridge case evaluations by set type (hard errors highlighted in bold)

| | Bullet Evaluations | | | | | |
|---|---|---|---|---|---|---|
| | ID | Inconclusive-A | Inconclusive-B | Inconclusive-C | Elimination | Total Sets |
| **Matching** | 76.6% | 9.04% | 8.90% | 2.56% | **2.92%** | 1405 |
| **Nonmatching** | 0.704% | 9.43% | 29.8% | 26.2% | 33.8% | 2842 |
| | Cartridge Case Evaluations | | | | | |
| | ID | Inconclusive-A | Inconclusive-B | Inconclusive-C | Elimination | Total Sets |
| **Matching** | 74.4% | 12.5% | 9.86% | 1.55% | **1.76%** | 1420 |
| **Nonmatching** | 0.917% | 6.24% | 22.5% | 21.9% | 48.5% | 2835 |

The numbers of examiners making each type of error are shown in Table 4. The false positive and false negative errors were made by a relatively small subset of the examiners as also reported in [7]. One hundred thirty nine of 173 examiners (80%) made no hard errors, either F-Pos or F-Neg when examining bullets; 137 made no hard errors of either kind when examining cartridge cases. In the first round of the study (Accuracy), hard errors were made by 34 of the 173 examiners when examining bullets; 36 of 173 for cartridge cases. Three participants made both kinds of errors. Analysis of the data collected in the first round of the study showed that the six most error-prone examiners account for 33 of the 112 errors - 29%, while 13 examiners account for almost half of all the hard errors (54 of 112 errors).

Table 4: Examiners making hard errors

| | Bullet Evaluations in the Accuracy Round of the Study | | | |
|---|---|---|---|---|
| | Zero False Negatives | One False Negative | Two or More False Negatives | Total Examiners |
| Zero False Positives | 139 | 17 | 7 | 163 |
| One False Positive | 3 | 1 | 1 | 5 |
| Two or More False Positives | 4 | 0 | 1 | 5 |
| Total Examiners | 146 | 18 | 9 | 173 |
| | Cartridge Case Evaluations in the Accuracy Round of the Study | | | |
| | Zero False Negatives | One False Negative | Two or More False Negatives | Examiners |
| Zero False Positives | 137 | 14 | 4 | 155 |
| One False Positive | 9 | 3 | 0 | 12 |
| Two or More False Positives | 6 | 0 | 0 | 6 |
| Total Examiners | 152 | 17 | 4 | 173 |

An obvious concern in this case is the possibility that error probabilities are different for individual examiners. If true, then regarding each comparison in the entire collection of examinations of matching bullet sets as having the same probability of being mistakenly labeled an Elimination (for example), is not an appropriate assumption. To examine this possibility, chi-square tests for independence were performed on tables of counts with 173 rows (one for each examiner), and 5 columns for examination results. For matching sets the proportions of Identification evaluations versus pooled Elimination and Inconclusive evaluations were compared, and for nonmatching sets the proportions of Elimination evaluations versus pooled Identification and Inconclusive evaluations were compared. Pooling of counts was used for these statistical tests because hard errors are relatively rare and, if maintained as a separate category, would result in many zero counts, which are problematic in chi-square tests, e.g. [30] For both matching and nonmatching sets, and for both bullets and cartridge cases, the hypothesis of independence was rejected ($p < 0.001$), strongly suggesting that the probabilities associated with each conclusion are not the same for each examiner. As a consequence, the most common methods of computing confidence intervals for proportions based on an assumption of equal probabilities for each evaluation category, e.g., the Clopper-Pearson intervals [31], are not appropriate.

A more appropriate procedure is based on an assumption that each examiner has an individual error probability, that these probabilities are adequately represented by a beta distribution, which is a flexible two-parameter probability distribution on the unit interval, across the population of examiners, and that the number of errors made by each examiner follows a binomial distribution characterized by that examiner's individual probability. Estimates and confidence intervals for the false positive error rate were also calculated using a beta-binomial model, as in the Baldwin study [7]; an example of its use in another application is given in [32]. Usual confidence intervals, in contrast, are based on an assumption that there is only one relevant binomial distribution, and that all examiners operate with the same error probability – an assumption our analysis strongly contradicts. Based on the beta-binomial model, maximum likelihood estimates and 95% confidence intervals for false positive and false negative error probabilities, integrated over all examiners, were calculated using the R statistics package, including the VGAM package [33,34]. The results are summarized in Table 5. The maximum likelihood estimates and confidence intervals are estimates of the mean of the examiner-specific error probabilities.

Table 5: Estimates and 95% confidence intervals for overall error probabilities, assuming different error probabilities for each examiner

| | **Bullet Comparisons** | | |
|---|---|---|---|
| | Point estimate | Lower 95% confidence limit | Upper 95% confidence limit |
| False Negative Probability | **2.87%** | 1.89% | 4.26% |

| False Positive Probability | **0.656%** | 0.305% | 1.42% |
|---|---|---|---|
| **Cartridge Case Comparisons** | | | |
| | Point estimate | Lower 95% confidence limit | Upper 95% confidence limit |
| False Negative Probability | **1.87%** | 1.16% | 2.99% |
| False Positive Probability | **0.933%** | 0.548% | 1.57% |

These confidence intervals should not be interpreted as bounding the error probabilities of any one examiner. Again, error probabilities of individual examiners are assumed to be different, and the data available for any one examiner are limited. A valid alternative explanation of the interval is that if many examiners were randomly selected from the population and individually asked to make a single determination for a (different) comparison set, the intervals specified would bound, with stated confidence, the overall proportion of errors made in this process.

It should also be noted that this method is not completely assumption-free (even though the assumptions are less restrictive than those on which the Clopper-Pearson intervals are based). Specifically, it is assumed without formal evidence that the beta distribution is appropriate for modeling the population of examiner-specific errors probabilities. The flexibility of the beta distribution family (i.e., the variety of shapes the distribution can take, controlled by its parameters) ensures that the methodology can be appropriate for a wide variety of situations. Because the examiner-specific error probabilities are not directly observable, and there is relatively limited information available on the accuracy of each examiner's determinations, it would be difficult to build a supportable case for a more appropriate distribution. Even if a different distribution were available, the beta distribution is certainly a more appropriate approximation than the single-value distribution assumed by the Clopper-Pearson approach.

Given the above provisos, the results of Table 5 should still be considered as approximate since the model of handgun and the positioning of known and questioned rounds in the firing sequence for a firearm also appear to affect error probabilities, and these considerations are not taken into account in this calculation; these effects will be discussed in a future paper. Still, differences among examiners are likely the greatest source of non-independence in the data, and the assumptions underlying the method used here are more appropriate than those upon which simpler methods are based.

# Discussion

**Accuracy**

In discussing the results of this study, comments will be directed initially in comparison to those reported in [7], which is cited frequently and was reviewed favorably by PCAST as fundamentally sound and statistically valid. In doing so, it is important to remember the distinct differences that exist between the experimental parameters of the two studies resulting in comparisons in the present study being more challenging (e.g., use of firearms that mark more poorly, steel vs. brass cartridge cases, presence of subclass characteristics). These differences are summarized in Table 6 below.

Table 6: Comparison of Baldwin et al. [7] vs. this study.

| **Design Attributes** | **Baldwin Study** | **Present Study** |
|---|---|---|
| Purpose | Accuracy | Accuracy<br>Repeatability<br>Reproducibility |
| Set Design | Open | Open |
| Specimen Examined | Cartridge cases | Cartridge cases,<br>Bullets |
| Ammunition Used | Brass Remington UMC<br>9 mm Pistol and Revolver<br>Cartridges | Copper + steel jacketed lead<br>Steel cartridge case<br>Wolf Polyformance 9 mm |
| Firearms Used | 25 Ruger SR9 9mm<br><br>Randomly acquired; not necessarily sequentially machined. | Cases: 10 Jimenez JA9 9mm<br>       27 Beretta M9A3 9mm<br>Bullets: 10 Ruger SR9c 9mm<br>       27 Beretta M9A3 9mm<br>(23 of 27 Berettas and all Rugers were sequentially machined). |

| Firing Sequence | Specimens were fired in groups of 100 and all comparisons were within the group | Specimens were fired in groups of 50. Specimens were divided into three ranges (Early, Middle, Late; EML) and comparisons were made for 9 EML combinations. |
|---|---|---|
| Number of Specimens Collected per Firearm | Cartridge cases: 800 | Cartridge cases: Jimenez 850<br>Beretta 700<br>Bullets:   Ruger 850<br>Beretta 700 |
| Number of Examiners | 218 | 173 |
| Comparisons in One Mailing | 15 Cartridge case comparisons (5 same-source firearms, 10 different-source firearms) | 15 Cartridge case comparisons (variable-3 to 7 same-source firearms, 8 to 12 different source firearms)<br><br>15 Bullet Comparisons (variable-3 to 7 same-source firearms, 8 to 12 different source firearms) |
| Single Comparison Set Makeup | 3 Knowns to 1 Questioned | 2 Knowns to 1 Questioned |
| Number of Case Comparisons | 3270 | 10,110 |
| Number of Bullet Comparisons | - | 10,020 |
| Error Rate (FP/FN): Cases | 0.939% / 0.367% | 0.933% / 1.87% |
| Error Rate (FP/FN): Bullets | - | 0.656% / 2.87% |
| ID Decision Basis Information Collected | No | Yes |

The error rates in this study are consistent with the overall error rates stated by Baldwin [7]. The Baldwin study did not examine bullets, so a comparison in that area between the two studies is not possible. Estimated error rates are higher in this study than those reported in a recent study that did examine bullets [24]. Results are also comparable to those of a more recent study [24] involving bullets fired from 30 consecutively machined Beretta barrels, which reported false positive rates of 0.08% (with 95% upper confidence limit up to 0.4%) and false negative error rates of 0.16%.

While the estimates of false positive rates reported here are comparable to those presented in the Baldwin study, the somewhat higher false negatives recorded herein are possibly due to greater difficulties when faced with the steel Wolf Polyformance cartridge cases rather than brass Remington UMC since many examiners commented that they felt brass provides better marks for Identification than steel. Anecdotally, the Jimenez firearm is known to generate gross marks with high occurrences of subclass characteristics both for breech face marks and firing pin impressions compared to higher cost-point firearms such as the Berettas (38). Thus, examinations of steel cartridge cases fired from a Jimenez firearm can be expected to present a more difficult comparison to an examiner than the specimens provided by Baldwin. The consecutively manufactured barrels and slides used in this study suggest another source of subclass characteristics. The consecutively machined barrels and the lack of a dropping breach in the firearms used in this study increase the difficulty level of comparisons. Examiners also observed that no contextual information was provided concerning the specimens, such as the possible time interval between the firing of the questioned specimen and the provided known specimens for comparison.

Various comments by examiners received during the course of the study attest to the difficulty of the comparisons, and some example comments are included to provide context. Concerning the type of firearm possibly leading to difficulties:

*"One of the major concerns is that they are presenting bullets with a high potential for subclass on some of those samples, and no mechanism for absolutely resolving that issue in the test, yet its resolution in a true laboratory setting would be as simple as doing a barrel cast and a quick visual exam…"*

*"In this research study, there were significant limitations on my ability to "evaluate the background" of the samples. Two questioned and one unknown sample. That's it. This study has really made me evaluate how "external information" influences my opinions; NOT regarding whether something is an IDENTIFICATION or not, but more so if two items should be ELIMINATED or INCONCLUSIVE."*

Various comments were received concerning the absence of firing order information and how it might affect an examiner's conclusions. For example:

*" … When we get test fires from a firearm, those test fires represent the condition of that gun at the time it was recovered. That becomes a baseline on which we can have absolute confidence. …   One of the first and most critical questions in any test such as this is whether the "known" samples are being*

*collected at the correct and relevant intervals with respect to the unknowns that are being generated. In normal case work TIME between the shooting event and the recovery of the firearm is a known."*

*"… when comparing these samples, even though the questioned item may have had absolutely no similarities with the known samples, I was very hesitant to eliminate it, even though I am very confident they were not fired in the same firearm. With a "single" unknown sample, there is no demonstrated repeatability of the patterns visible on this item. There is no way to determine or even estimate if there was 1 shot between the questioned and unknown or 5000 shots…."*

The concentration of the errors by a relatively small number of examiners, as was seen in Baldwin, was again noted. Examination of the data using chi-square tests for independence show that the numbers cited above cannot be applied equally to all examiners; most examiners will perform better than the percentages cited above while a few will perform more poorly. Point estimates and confidence intervals were calculated under the assumption (supported by our analysis) that examiners have different error probabilities, and that the collection of examiner-specific probabilities can be represented by a beta distribution. The 95% confidence intervals presented in Table 5 represent the overall error rate that would be expected for a randomly selected examiner when asked to evaluate a randomly selected cartridge case or bullet set. The maximum likelihood estimates of error rates listed in this table should be interpreted the same way; they, rather than the overall simple proportions cited above in Table 5, should be regarded as the definitive error rate estimates derived in this study.

*Inconclusive Determinations*

A clear difference exists between what are termed hard errors and inconclusive decisions. Forensic examination must be regarded as at least a two-step process. The first step is an evaluation of the class characteristics. If they are congruent, the second step involves comparing the quality and quantity of microscopic correspondence. For many reasons, fired bullets and cartridge cases do not always carry marks sufficient to support a definitive conclusion of Identification or Elimination.

The production of individual characteristic marks in any given firing is a random event that is subject to limiting factors, such as, variations in metallic properties of the substrate, changes in a tool's working surfaces due to wear, corrosion, obturation, fouling, and damage. When there is an inadequate reproduction for a toolmark, the quality and/or quantity of individual characteristics available for comparison may be insufficient to conclude an identification or elimination. In this situation, the appropriate recourse for an examiner is a decision of inconclusive.

*Comparisons to Other Forensic Fields*

Accuracy studies have been conducted in other areas of forensic comparisons including fingerprint analysis [-1,35] and footwear analysis [36]. Straightforward comparison of the results of this paper to these other fields is difficult considering that each has its own rating system, methodology, and criteria for reaching a conclusion, and any attempt to use a "one-size-fits-all" analysis system must be approached with caution. One possibly useful analysis is to use the data of Table 2 to calculate values for overall sensitivity and specificity of the examinations.

Table 7 gives commonly reported indices of sensitivity and specificity computed from our data. Note that these are simple ratios of bulk counts, and in the light of our discussion concerning unequal error probabilities among examiners, are intended primarily for comparison to other studies rather than as preferred estimates of meaningful underlying parameters. Sensitivity is defined as the number of Identification evaluations reported divided by the number of total known matches based on ground truth. It is a measure of the study participant's ability to identify a match between two specimens when they are from the same source. Similarly, specificity is the number of Elimination evaluations reported divided by the number of total nonmatches based on the ground truth. Note that sensitivity would be one minus the false negative error rate, and specificity would be one minus the false positive rate, <u>if</u> Inconclusive evaluations were not allowed (and not accommodating variation in individual examiner error rates):

Table 7: Computed indices of sensitivity and specificity.

| Bullets Comparisons | | |
|---|---|---|
| Factor | Calculation | Percent |
| Sensitivity [Identifications] | 1076/1405 | 76.6% |
| Specificity [Eliminations] | 961/2842 | 33.8% |
| **Cartridge Case Comparisons** | | |
| Sensitivity [Identifications] | 1056/1420 | 74.4% |
| Specificity [Eliminations] | 1375/2835 | 48.5% |

Higher values exist for sensitivity while lower values are obtained for specificity. While the sensitivity values are on the order of those noted for fingerprint analysis [35], the lower specificity value indicates that it was more difficult for firearms examiners to justify an Exclusion decision. Some explanation for this is straightforward. Examiners commented that without having the actual firearm in hand to test, they found it difficult to render an Exclusion, particularly when there was no information given as to the time gap between the collection of the unknown and the collection of the known exemplars. Indeed, in many cases particpants' laboratory policy is not to render an Exclusion decision in the absence of such information. Therefore, the low specificity values should not be surprising.

In light of the possibility that administrative policy or other factors may have an effect, specificity values were recalculated by pooling the number of Inconclusive-C responses with Elimination. To provide a comparison as to the effect of pooling sensitivity was also recalculated pooling Identification with Inconclusive-A responses. These values are shown in Table 8. Calculated specificity must improve when Inconclusive-C is combined with Elimination, as must sensitivity when Inconclusive-A is combined with Identification, but the effect is much more pronounced in specificity.

Table 8: Sensitivity and Specificity values when pooling is included.

| **Bullet Comparisons** | | |
|---|---|---|
| Factor | Calculation | Percent |
| Sensitivity (TPR) [Identifications + Inc A] | 1203/1405 | 85.6% |
| Specificity (TNR) [Eliminations + Inc C] | 1706/2842 | 60.0% |
| **Cartridge Case Comparisons** | | |
| Sensitivity (TPR) [Identifications + Inc A] | 1233/1420 | 86.8% |
| Specificity (TNR) [Eliminations + Inc C] | 1995/2835 | 70.4% |

**Other Considerations**

The constraints imposed by this study, i.e., absence of contextual information concerning the specimens that might affect the conclusion of the examiner, were frequently mentioned by participants as being a challenge. Comments received from examiners during the study were generally that the study parameters did not allow them to follow normal procedures that would have better enabled them to arrive at either an Identification or Elimination.  Not having access to the actual firearm and not knowing the number of firings between specimens provided for examination were common observations that examiners stated would cause them to be more cautious than they ordinarily would have been when declaring Eliminations.

## Summary and Conclusions

The experimental results obtained in this study are consistent with those of previous studies [7, 10] [refs in [9].  The following conclusions can be drawn:

1. Estimates for overall false positive and false negative error probabilities were calculated as 0.656% and 2.87% for bullets and 0.933% and 1.87%, for cartridge cases, respectively. The 95% confidence intervals for false positives and false negatives are (0.305%,1.42%) and (1.89%, 4.26%), respectively, for bullets.  Similarly, for cartridge cases the 95% confidence intervals are (0.548%, 1.57%) and (1.16%, 2.99%) for false positives and false negatives, respectively.

2. The majority of errors were made by a limited number of examiners. For example,

Thirteen examiners account for almost half of all the hard errors (54 of 112); a subset of these 13 are the six most error-prone examiners, who accounted for almost 30% (33 of the 112) of the total errors.

3. Comments from examiners indicated that the "black box" nature of this study presented challenges that might normally be overcome in an actual laboratory setting. Not having access to the firearm to make additional comparison specimens, not knowing the time between firings, and the type of firearms and ammunition used all contributed to making this study a difficult test of examiner skills.

4. Calculated specificity increases more when Inconclusive-C is combined with Elimination, than calculated sensitivity increases when Inconclusive-A is combined with Identification. This suggests that examiners, either by inclination or lab policy, may be reluctant to declare Eliminations without having more contextual information and/or the actual firearm in their possession.

# **Acknowledgments**

This is manuscript 21-29 of the FBI Laboratory. Names of commercial manufacturers are provided for identification purposes only, and inclusion does not imply endorsement of the manufacturer or its product or services by the FBI. Any reproduction, or other use of these presentation materials without the express written consent of the FBI is prohibited. The views are those of the author(s) and do not necessarily reflect the official policy or position of the FBI or the U.S. Government.  This study was funded by the Federal Bureau of Investigation (FBI) through contract with the Ames Laboratory, U.S. Department of Energy, under Interagency Agreement DJF-15-1200-0000667, with supplemental funding from the National Institute of Justice. The authors wish to acknowledge Dr. David Baldwin who was responsible for conducting an earlier study and his efforts in the early stages of this study. The authors also acknowledge useful discussions with Dr. Michael Smith and Ms. Heather Seubert.  Special thanks are due to Ms. Jennifer Stephenson for assisting in the collection of the specimens used in this study. We appreciate the graciousness of management and staff at the Beretta, Ruger, and Jimenez firearms factories who provided production details, extended observations on the factory floor, unrestricted conversations with staff, and permitted still and video photography of their operations.  This work was

38. Private communications between the research team and numerous examiners that participated in the study.